\documentclass[10pt,aps,prapplied,
  superscriptaddress,
  twocolumn,
  floatfix
]{revtex4-2}

\usepackage{scrbase}

\usepackage[english]{babel}
\usepackage[caption=false]{subfig}
\usepackage{upgreek,graphicx,dcolumn,contour,xspace,lipsum}
\usepackage{amsmath,amssymb,physics}
\usepackage{siunitx}
\usepackage{chemmacros}
\usepackage[version=4,arrows=pgf]{mhchem}
\usepackage{comment}

\usepackage{amsmath,amssymb,physics}
\usepackage{siunitx}
\usepackage{chemmacros}
\usepackage[version=4,arrows=pgf]{mhchem}

\usepackage{todonotes}
\usepackage[normalem]{ulem}

\renewcommand{\selectlanguage}[1]{}

\setuptodonotes{inline,color=green!50}

\usepackage{etoolbox}
\robustify{\subref}

\usepackage[hidelinks]{hyperref}

\usepackage[commandnameprefix=always]{changes}
\usepackage{cleveref} 

\graphicspath{{./}{./img/}{./img/tikz/}}

\newcommand{\approxprop}{\mathrel{\vcenter{
  \offinterlineskip\halign{\hfil$##$\cr
    \propto\cr\noalign{\kern2pt}\sim\cr\noalign{\kern-2pt}}}}}

\newcommand\emx[1]{\ensuremath{#1}\xspace}

\newcommand\freq{\emx{f}}

\newcommand\stress{\emx{\sigma}}

\newcommand\modeno{\emx{n}}
\newcommand\youngs{\emx{E}}
\newcommand\thick{\emx{h}}
\newcommand\length{\emx{L}}
\newcommand\dens{\emx{\rho}}

\begin{document}


\title{
All-dry processing of 3C-SiC nanomechanical string resonators for extreme aspect ratios and high intrinsic quality factor}

\author{Felix David}
\affiliation{Department of Electrical Engineering, 
             School of Computation, Information and Technology,
             Technical University of Munich,
             85748 Garching, Germany}
\author{Philipp Bredol}
\email{philipp.bredol@tum.de}
\affiliation{Department of Electrical Engineering, 
             School of Computation, Information and Technology,
             Technical University of Munich,
             85748 Garching, Germany}
\author{Yannick S. Kla\ss}
\affiliation{Department of Electrical Engineering, 
             School of Computation, Information and Technology,
             Technical University of Munich,
             85748 Garching, Germany}
\author{Eva M. Weig}
\email{eva.weig@tum.de}
\affiliation{Department of Electrical Engineering, 
             School of Computation, Information and Technology,
             Technical University of Munich,
             85748 Garching, Germany}
\affiliation{Munich Center for Quantum Science and 
             Technology (MCQST), 80799 Munich, Germany}
\affiliation{TUM Center for Quantum Engineering (ZQE), 
             85748 Garching, Germany}

\date{\today}


\begin{abstract}
Conventional fabrication of suspended nanomechanical resonators typically relies on wet-chemical process steps and critical point drying, which can compromise sample yield and cleanliness. Here, we present an all-dry fabrication process for strongly stressed 3C-\ce{SiC} nanomechanical string resonators that entirely avoids wet-chemical etching and cleaning. Using a negative-tone electron-beam resist as an etch mask and a three-step reactive-ion etching process for both structuring and release, we achieve high fabrication yield, clean suspended structures, and extreme aspect ratios of $\num{8,500}$. We perform full mechanical characterization of the resulting doubly-clamped nanostring resonators, and establish a benchmark intrinsic quality factor for dissipation-diluted 3C-\ce{SiC} of $Q_{\text{intr}}=\num{4,200}$.

\end{abstract}

\maketitle


Crystalline \ce{SiC} is a compelling material platform for nanomechanics, combining a wide electronic bandgap and high thermal conductivity with excellent mechanical robustness and a high Young's modulus and yield strength. It is compatible with silicon substrates and conventional silicon process technology, allowing integration into established fabrication workflows, and is amenable to wafer-scale, industrial growth. Crucially for nanomechanical applications, the cubic polytype 3C-\ce{SiC} can be grown as thin films with strong intrinsic tensile pre-stress.

In nanomechanical resonators, tensile pre-stress enables dissipation dilution~\cite{gonzales,unterreithmeier,Yu-diss-dil}, a mechanism that has transformed the field of nanomechanics over the past decade, driving quality factors to unprecedented levels by enabling further concepts such as soft clamping~\cite{tsaturyan-soft-clamping} and strain engineering~\cite{ghadimi-strain} that build directly on the dilution principle~\cite{sementilli_review, fedorov-dis-dill}. Dissipation dilution has been most extensively studied in amorphous, LPCVD-grown \ce{Si3N4}~\cite{verbridge-sin,Vill-schm-diss_dil}, but crystalline, stressed materials are increasingly attracting attention, including Si~\cite{beccari-10bill}, SiC~\cite{romero-diss-dil-sic}, InGaP~\cite{Manheshwar-InGaP}, and AlN~\cite{ciers-AlN} for their potentially higher intrinsic quality factors, as well as their specific material properties such as, e.g. piezoelectricity or integrability with spin defects.

Maximizing dissipation dilution requires simultaneously high tensile stress and a large aspect ratio, besides a high intrinsic quality factor. 
Achieving large aspect ratios, however, is typically fabrication-limited: Releasing nanostructures from the substrate conventionally relies on wet-chemical etching, where surface tension during drying frequently causes stiction or collapse, preventing the release of extremely high-aspect ratio structures or otherwise significantly reducing the fabrication yield.

Here, we present an all-dry fabrication process for 3C-\ce{SiC} nanomechanical resonators that entirely avoids wet-chemical etching and cleaning, instead employing a three-step plasma process in an inductively coupled plasma (ICP) reactive-ion etcher. This process yields higher fabrication yield and cleaner structures, and enables extreme aspect ratios -- all within a simple, single-tool process. Nanomechanical characterization further reveals a benchmark intrinsic quality factor for 3C-\ce{SiC}.

\section{All-dry processing of nanomechanical string resonators}

\begin{figure}
    \centering
    \includegraphics[scale=0.55]{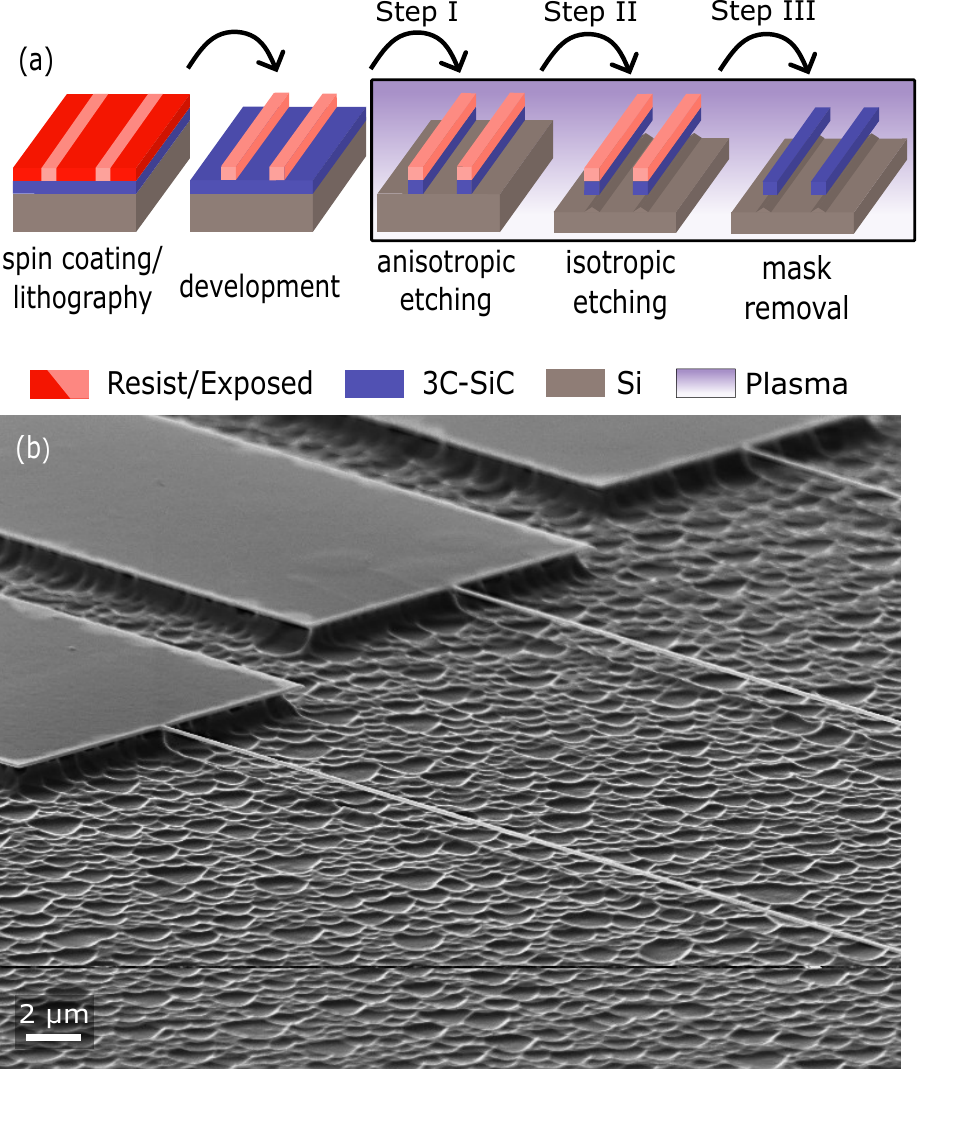}
    \caption{Dry-processing of 3C-SiC nanomechanical resonators. (a) Schematic process flow. (b) Scanning electron micrograph of a typical device, depicting three string resonators with a width (thickness) of \SI{250}{nm} (\SI{115}{nm}), and a length of \SI{110}{\micro\metre} (front), \SI{100}{\micro\metre} (middle) and \SI{90}{\micro\metre} (back).} 
    \label{Fab_result}
\end{figure}

The nanomechanical resonators are processed from a \SI{115}{nm}-thick, crystalline 3C-\ce{SiC} thin film grown on a (111) silicon substrate. The lattice mismatch of approx. \SI{20}{\%} between the SiC film and the substrate \cite{chen-epi-sic-si} induces strong tensile prestress in the \ce{SiC} film. Neither the film nor the substrate is intentionally doped. The sample preparation is schematically described in Fig.~\ref{Fab_result}(a). The wafer is diced into \SI{5}{mm} x \SI{5}{mm} chips which subsequently undergo ultrasonic cleaning in acetone followed by a rinse in isopropanol. We then spin-coat a \SI{450}{nm}-thick layer of \textit{AR-N 7520.17 new} resist (Allresist) and expose it with doubly-clamped nanomechanical string resonator structures of varying length using an Elionix ELS-BODEN100 electron-beam lithography system at an acceleration voltage of \SI{100}{\kV}. After development, the resist pattern is used as an etch mask to subsequently define and release the string resonators before removing the resist mask.

To this end, we use inductively coupled plasma reactive ion etching (ICP-RIE) in a Cobra 80 etching system (Oxford Instruments).
Our etching process consists of a three-step plasma-etching recipe to achieve freestanding nanomechanical structures: (I) an anisotropic \ce{SiC} etch~\cite{Kelner1987,Pan1990,Jiang2003,forster2004}, (II) an isotropic \ce{Si} etch, and (III) an oxygen-plasma resist removal. 
The first anisotropic step cuts through the \ce{SiC} layer using a \ce{SF6}/\ce{Ar} mixture. To achieve vertical sidewalls for our nanomechanical devices, gas flows (\SI{2}{sccm} \ce{SF6}, \SI{4}{sccm} \ce{Ar}), pressure (\SI{25}{mTorr}), and HF/ICP powers (\SI{20}{W}/\SI{200}{W}) are adjusted to obtain a high DC bias of around \SI{100}{V}, resulting in an etch rate of \SI{0.96}{\frac{nm}{s}}. This is sufficiently fast to etch thin-film crystalline \ce{SiC} with a resist mask. Note that the etching rate of resist is about twice of \ce{SiC}, so with the \SI{450}{nm} resist mask we can cut through more than \SI{200}{nm} of \ce{SiC}. The second, isotropic step uses the same gas mixture of \ce{SF6} and \ce{Ar}, with higher flow rates of \SI{5}{sccm} for \ce{Ar} and \SI{30}{sccm} for \ce{SF6}. We reduce the HF power to \SI{20}{W} and disable the ICP, which yields a DC bias of around \SI{50}{V} and results in isotropic \ce{Si} etching. Similar recipes for isotropic \ce{Si} etching have been reported in the literature \cite{si-etch,si-etch-chen}, but have not previously been employed to fabricate freely suspended nanomechanical structures. In the third step, we remove the remaining resist mask with oxygen plasma with \SI{50}{sccm} for \ce{O2} and a exposure time of \SI{6}{min}. Figure~\ref{Fab_result}(b) depicts a scanning electron micrograph of dry-processed nanostring resonators. Each of these resonators is \SI{250}{nm} wide and \SI{115}{nm} thick. Clearly, the \ce{SiC} nanostring is freely suspeneded above the Si substrate. Notice that the rough substrate surface results from etching the triangular voids at the 3C-\ce{SiC}-Si interface \cite{book-sic}. Compared to established process flows, where an anisotropic ICP etch is combined with an isotropic wet etch and/or wet-chemical etch mask removal, our all-dry process does not require critical point drying. This avoids the particle contamination that critical point drying frequently introduces, and yields substantially cleaner samples. The dry process further significantly improves the fabrication yield.
Finally, it enables the realization of nanostrings with extreme aspect ratios and ultralong string geometries that are not achievable with wet-etching-based approaches. Furthermore, it also allows to etch relatively flat, which simplyfies interferometric measurements. In our process, we do not etch deeper than \SI{1.5}{\micro\metre}.

\section{Mechanical characterization}
In the following, we explore the mechanical properties of a sample hosting $16$ sets of doubly clamped, tensile-stressed nanostring resonators. Each set comprises 10 string resonators with lengths ranging from \SI{20}{\micro\metre} to \SI{110}{\micro\metre} in increments of \SI{10}{\micro\metre}, all with a thickness of \SI{115}{nm} and a width of \SI{250}{nm} (see Fig.~\ref{Fab_result}(b)). We use optical interferometric detection and mechancial actuation with a piezo transducer to measure the reponse of the resonators, as described in elsewhere~\cite{helium-implantation}.  All measurements are performed under vacuum conditions at approx. \SI[parse-numbers=false]{10^{-3}}{\milli\bar} and at room temperature. The frequency response of all nanostrings is measured using a vector network analyzer. We extract the resonance frequencies of a set of out-of-plance flexural eigenmodes for all investigated resonator lengths.

\begin{figure*}
    \centering
    \includegraphics[width=1\linewidth]{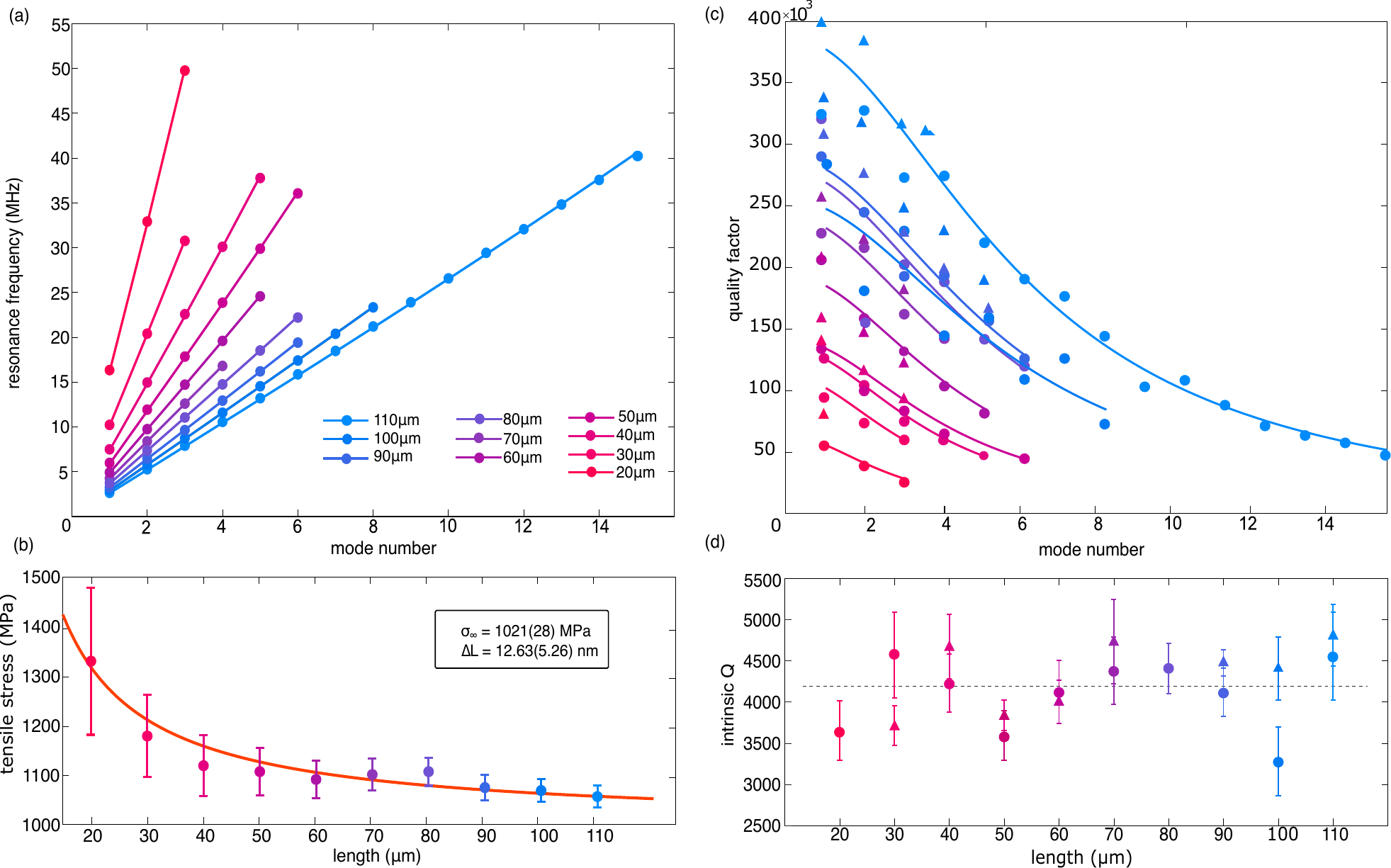}
    \caption{a) Resonance frequency $f$ as a function of mode number $n$ for string lengths between \SI{110}{\micro\meter} and \SI{20}{\micro\meter}, with fits of the Euler-Bernoulli model Eq.~\ref{Eq-EBB} (solid lines). b) Resulting tensile stress $\sigma$ as a function of string length $L$, including fit of the expected $1/L$ behavior (solid line). c) Mechanical quality factor $Q$ as a function of mode number $n$ extracted from spectral (circles) and ringdown (triangles) measurements for different string lengths. Solid lines indicate fits of the spectral data with Eq.~\ref{Eq-DD}. Fits of ringdown data are not shown for clarity. d) Resulting intrinsic quality factor obtained from spectral (circles) and ringdown (triangles) data for each length.}
    \label{EBB}
\end{figure*}
First, we analyze the obtained eigenfrequencies. Figure~\ref{EBB}(a) depicts all recorded frequencies from a set of doubly clamped string resonators, with lengths between $\SI{20}{\micro\meter}$ and $\SI{110}{\micro\meter}$. The recorded frequencies span from \SI{2}{MHz} to \SI{50}{MHz}, with up to $16$ eigenmodes per resonator. We fit the data with the eigenfrequency relation from Euler-Bernoulli beam theory under simply-supported boundary conditions,
\begin{align}
  \freq(\modeno)=\frac{\modeno^2\pi}{2\length^2}\sqrt{\frac{\youngs\thick^2}{12\dens}}\sqrt{1+\frac{12\stress\length^2}{\modeno^2\pi^2\youngs\thick^2}},
  \label{Eq-EBB}
\end{align}
where $n$ is the mode number, $L$ the (nominal) length, assuming an error of 
$\pm\SI{1}{\micro\metre}$, $h=\SI{115\pm5}{nm}$ the thickness, 
$\rho = \SI{3200\pm10}{\kilogram\per\cubic\metre}$ describes the density of \ce{SiC}~\cite{book-sic}, $E =\SI{400\pm38}{GPa}$ the Young's modulus~\cite{klass-youngs}, $\sigma$ the tensile pre-stress. This leaves $\sigma$ as the only free fit parameter. The resulting tensile stress for each length, shown in Fig.~\ref{EBB}(b), is found to vary between \SI{1,100}{MPa} for the longest and \SI{1,350}{MPa} for the shortest string, confirming the established  $1/L$ scaling~\cite{bueckle-length-dep}.
Second, we analyze the mechanical quality factor of the investigated strings. The mechanical quality factor is determined from the linewidth of the response curve by fitting a Lorentzian. For modes with a high quality factor, we additionally perform ringdown measurements to exclude systematically underestimating the results as a result of spectral broadening~\cite{maillet}. Quality factors obtained from both methods are depicted in Fig.~\ref{EBB}(c) as solid circles and triangles. The fundamental mode exhibits the highest $Q$ for every string, with Q decreasing monotonically for higher harmonics. This fundamental-mode $Q$ is in turn highest for the longest string, reaching a maximum value of \num{400,000} and decreases with decreasing string length, with shortest string remaining below \num{100,000}. Spectral and ringdown measurements agree well with each other. The quality factor of a pre-stressed nanomechanical resonator is described by the dissipation dilution model~\cite{gonzales,unterreithmeier,Yu-diss-dil,sementilli_review,fedorov-dis-dill}. It provides an analytic description of the stress-enhanced stored energy resulting from the lossless potential induced by the tension (and geometric nonlinearity) of the string~\cite{fedorov-dis-dill}, under the assumption that extrinsic loss channels are negligible, such that the measured $Q$ is limited only by intrinsic material dissipation.
For a one-dimensional string it yields
\begin{align}
Q = Q_{\text{intr}}\left(\frac{2h}{\length}\sqrt{\frac{\youngs}{12\stress}}+\modeno^2\pi^2\left(\frac{h}{\length}\sqrt{\frac{\youngs}{12\stress}}\right)^2\right)^{-1},
\label{Eq-DD}
\end{align}
where $Q_{\text{intr}}$ denotes the intrinsic quality factor. We use the same values for the density $\rho$ and for the Young's modulus $E$ as in the eigenfrequency analysis, along with the values of the length-dependent tensile stress $\sigma$ determined from the  Euler-Bernoulli fit of Eq.~(\ref{Eq-EBB}). Fits of Eq.~\ref{Eq-DD} to the data obtained using the spectral linewidth (circles) with $Q_{\text{intr}}$ as the only free fit parameter are included in Fig.~\ref{EBB}(c) as solid lines. The resulting intrinsic quality factor for each string length is plotted in Fig.~\ref{EBB}(d) as a circle. Intrinsic quality factors obtained from fitting the ringdown data are included as triangles. We find that $Q_{\text{intr}}$ is approximately constant across all string lengths, with a dashed line indicating the average value of \num{4,200(440)}. This value exceeds the intrinsic quality factor we obtain from conventionally processed 3C-\ce{SiC} resonators (combination of dry and wet etching, with a metallic hard mask), for which we previously found $Q_{\text{intr}} =\num{2,900(330)}$~\cite{Klass_diss}. Comparable values have been reported for as-processed 3C-SiC with similar thickness \cite{romero-diss-dil-sic}. Drawing on detailed analysis of stressed \ce{Si3N4} resonators~\cite{Vill-schm-diss_dil}, we assume that $Q_{\text{intr}}$ is surface-loss limited in that thickness regime. Only for considerably thicker layers approaching \SI{1}{\um}, higher $Q_{\text{intr}}\approx 8,000$ were reported \cite{garofal-new-Qintr}, reflecting the decreasing influence of surface losses~\cite{Vill-schm-diss_dil}.

\section{Extreme Aspect Ratios}
\begin{figure}
    \centering
    \includegraphics[width=1\columnwidth]{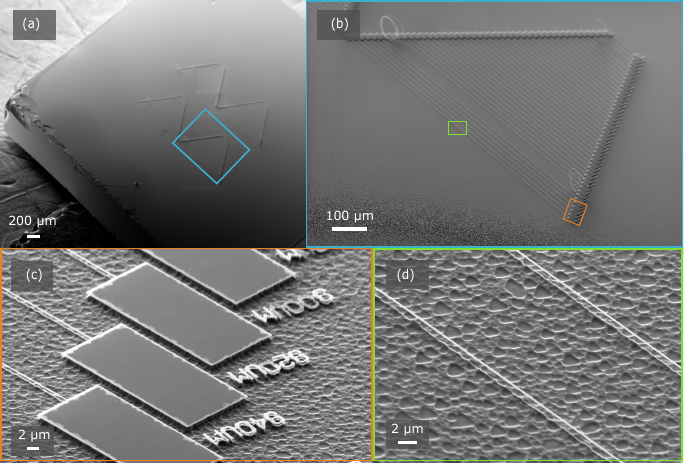}
        \caption{Scanning electron micrographs of dry-processed extreme-aspect ratio nanostrings.  (a) Overview of the $5x5$\,mm$^2$ chip, revealing four sets of string resonators of variable length. (b) close-up of one set of string resonators. Length varies between \SI{100}{\micro\metre} and \SI{940}{\micro\metre}. (c) Magnified view of the clamping region. (d) Magnified view of the center of two of the string resonators.}
    \label{SEM}
\end{figure}

Finally, we highlight the extreme aspect ratios enabled by the presented all-dry fabrication process. Figure~\ref{SEM} shows a second sample, featuring four series of ultralong doubly clamped nanostrings on a \SI[parse-numbers=false]{5x5}{\milli\meter} silicon chip. An overview of the complete sample is depicted in Fig.~\ref{SEM}(a), while Fig.~\ref{SEM}(b) shows one of the series, hosting string resonators of lengths between \SI{100}{\micro\metre} and \SI{940}{\micro\metre} in increments of \SI{20}{\micro\metre}. These string resonators are also \SI{115}{nm} thick and \SI{250}{nm} wide. Further close-ups are presented in Fig.~\ref{SEM}(c) and (d), clearly demonstrating that the strings have been fully released. The aspect ratio of the longest string amounts to $L/h \approx \num{8,500}$, promising to increase the dilution factor of the fundamental mode by about an order of magnitude. Mechanical quality factors of these ultralong strings are presently limited by gas damping; reducing the chamber pressure to ultrahigh-vacuum conditions will be required to obtain meaningful quality factor measurements.

\section{Conclusion}
In conclusion, we developed an all-dry fabrication process for nanomechanical string resonators from a strongly pre-stressed 3C-SiC thin film, avoiding wet-chemical release or cleaning steps entirely. Eliminating wet chemistry and critical point drying yields substantially cleaner samples and higher fabrication yield, and further enables reliable fabrication of string resonators with extreme aspect ratios not achievable with wet-etching-based approaches.
We performed a full mechanical characterization of these resonators which includes the resonance frequencies and quality factors of the fundamental and a large number of higher-harmonic out-of-plane flexural modes. The eigenfrequencies allow to extract the length-dependent tensile stress, which in turn is used to obtain the intrinsic quality factor of each string from the dissipation dilution model. We find average $Q_{\text{{intr}}}=\num{4,200(440)}$,
which exceeds the highest intrinsic quality factor reported for dissipation-diluted 3C-SiC by more than $50\%$. Beyond these measured devices, our process yields string resonators with lengths approaching a millimeter and aspect ratios of up to \num{8,500}. Such long, high-aspect-ratio strings are a key building block for advanced, ultra-high $Q$ resonator geometries~\cite{romero-diss-dil-sic, Klass_diss} including trampoline resonators~\cite{Reinhardt-trampoline}, spiderweb resonators~\cite{shin-spider-web}, soft-clamped strings~\cite{ghadimi-strain}, perimeter modes\cite{bereyhi-poly-3bill}, and binary-tree resonators~\cite{beccari-10bill}, and may thus prove valuable for ultra-sensitive force sensing and quantum optomechanics at room temperature.

\begin{acknowledgements}
We gratefully acknowledge financial support from the Deutsche Forschungsgemeinschaft (DFG, German Research Foundation) through Project-IDs No. WE 4721/1-1 and No.425217212-SFB 1432, as well as under Germany’s Excellence Strategy—EXC-2111—390814868. The research is further supported by the Bavarian state government with funds from the Hightech Agenda Bavaria.
We further thank the TUM Center for Nanotechnology and Nanomaterials (ZNN) and TUM Central Electronics and Information Technology Laboratory (ZEITlab) for shared use of their cleanroom facilities.
\end{acknowledgements}

\section*{Data Availability Statement}
The data that support the findings of this study are
openly available in 

%


\bibliography{main}

\end{document}